\documentstyle[aps,twocolumn,epsfig]{revtex}

\title{STABILITY AND HALO FORMATION \\ IN AXISYMMETRIC INTENSE BEAMS}
\author{Robert L.~Gluckstern,
        University of Maryland, College Park, MD 20742, and \\
        Sergey S.~Kurennoy, 
        LANSCE-1, LANL, MS H808, Los Alamos, NM 87545}

\begin{document}
\maketitle

\begin{abstract}
Beam stability and halo formation in high-intensity axi\-symmetric 2D 
beams in a uniform focusing channel are analyzed using particle-in-cell 
simulations. The tune depression - mismatch space is explored for
the uniform (KV) distribution of the particle transverse-phase-space 
density, as well as for more realistic ones (in particular, the 
water-bag distribution), to determine the stability limits and halo 
parameters. The numerical results show an agreement with predictions 
of the analytical model for halo formation \cite{RLG94}.
\end{abstract}

\section{Introduction}

There is an increasing interest in high-current applications of
ion linacs, such as the transformation of radioactive waste, the 
production of tritium, and fusion drivers. High currents of the 
order of 100~mA restrict beam losses below 1~ppm. 
Thorough studies are necessary to understand mechanisms of 
intense-beam losses, in particular, beam instabilities 
and halo formation.

Most of the theoretical efforts so far have concentrated on the 
Kap\-chin\-sky-Vladimirsky (KV) distribution of particles in 
transverse phase space \cite{KV}. The KV beam density is uniform 
so that space-charge forces inside the beam are linear. It allows 
an analytical investigation and results are used to predict the 
behavior of real beams. 
On the other hand, it is recognized that the KV model, 
in which all particles have the same transverse energy, is not a 
realistic beam distribution, e.g.\ \cite{Okam}. The present paper 
compares the KV beam with other, nonlinear particle-density 
distributions, which can serve as better models for real beams.

\section{Analytical Consideration}

We study a continuous axisymmetric ion beam in a uniform 
focusing channel, with longitudinal velocity $v_z=\beta c$. 
The Hamiltonian of the transverse motion ($v_\bot \ll v_z$) is
\begin{equation}
 H(r,s) = s^2/2 + k_0^2r^2/2 + q \Phi(r)/
  (m \gamma^3 \beta^2 c^2) \ , \label{H}
\end{equation}
where $m$ and $q$ are ion mass and charge, $k_0$ is the 
focusing strength of the channel, $\gamma = (1-\beta^2)^{-1/2}$, 
$r=\sqrt{x^2+y^2}$ is the distance from the $z$-axis in the 
transverse plane, and $s=\sqrt{x'^2+y'^2}$ 
($x'=\dot{x}/\beta c$, $y'=\dot{y}/\beta c$) is the dimensionless
transverse velocity. The electric potential $\Phi(r)$ must satisfy
the Poisson equation
\begin{equation}
 \nabla^2 \Phi(r) = - (q/\varepsilon_0) \int\!\!\int d \vec{s}
  f(r,s) \ , \label{pois}
\end{equation}
where $f(x,y,x',y')=f(r,s)$ is the distribution function in the 
transverse non-relativistic 4-D phase space.  The integral on the 
RHS is the particle density $d(r)$.

Since the Hamiltonian (\ref{H}) is an integral of motion, any 
distribution function of the form $f(r,s)=f(H(r,s))$ is a 
stationary distribution. We consider a specific set of stationary 
distributions for which the beam has a sharp edge (for all ions
$r \le r_{max}=a$), namely, 
\begin{equation}
  f_n(H) = \left \{ \begin{array}{cc}
       N_n n (H_0-H)^{n-1} &  \mbox{for } H \leq H_0 \ ,  \\
                   0 &  \mbox{for } H > H_0 \ .  \label{fn}
                \end{array} \right.
\end{equation}
The normalization constants $N_n$ are chosen to satisfy
$2\pi \int_0^a r dr d(r) = I$, where $d(r)$ is the particle density, 
and $I$ is the beam current.
The set includes the KV distribution, $f_0=N_0 \delta(H_0-H)$, 
as a formal limit of $n \to 0$, 
as well as the waterbag (WB) distribution, $f_1=N_1 \theta(H_0-H)$, 
where $\theta(x)$ is the step-function. For a detailed discussion of 
these two specific examples see \cite{MR}.

We introduce the function $G(r) = H_0 - H(r,s) + s^2/2$,
because the density can be expressed from (\ref{fn}) as 
$d(r)= 2\pi N_n G^n(r)$. Physically, this function gives the 
maximal transverse velocity for a given radius,
$s_{max}(r)=\sqrt{2G(r)}$, and defines the boundary
in the phase space $(r,s)$.
It allows us to rewrite Eq.~(\ref{pois}) as 
\begin{equation}
 \left [ r G'(r) \right ]'/r - \lambda^2 G^n(r) = 
    - 2 k_0^2 \ , \label{Gsc}
\end{equation}
with boundary conditions $G(a)=0$, and $G(0)\equiv G_0$ is finite. 
Here the parameter $\lambda^2 = K/\left[\int_0^a r dr G^n(r)\right]$, 
where $K=2I/\left(I_0\beta^3\gamma^3\right)$ is the beam perveance, 
and $I_0=4\pi \varepsilon_0 mc^3/q$ is a constant. 
Particular solutions to (\ref{Gsc}) are easy to find for 
$n=0$ (KV) and $n=1$ (WB) \cite{MR}. For $n \ge 2$ a numerical 
solution is required.

To compare different transverse distributions on a common basis, 
we consider rms-equivalent beams which have the same perveance $K$, 
rms radius, and rms emittance $\tilde{\mathcal{E}}$. 
To characterize the space-charge strength, one introduces an 
equivalent (or rms) tune depression
\begin{equation}
 \eta = \sqrt {\, 1- K/(4k_0^2x^2_{rms})} \ ,          \label{eta}
\end{equation}
which reduces to the usual one
for the KV beam.
 For numerical simulations we use dimensionless variables:  
$\hat{z}=k_0 z$, and $\hat{x}=x\sqrt{k_0/\mathcal{E}}$, where 
${\mathcal{E}} = a'a$. 
In normalized variables the beam matched radius is 
$ \hat{a} = \sqrt{(C_{\mathcal{E}}/C_2)/\eta}$,
where $C_{\mathcal{E}}=\tilde{\mathcal{E}}/
 \mathcal{E}$ and $C_2=x^2_{rms}/a^2$. For the KV case,
$C_{\mathcal{E}}=C_2=1/4$, so that $\hat{a} = 1/\sqrt{\eta}$.
 The ``hats'' are omitted below to simplify notation.

\section{Numerical Simulations}

We use particle-in-cell simulations to study beam stability
and halo formation in the presence of instabilities. A 
leap-frog integration is applied to trace the time evolution
for a given initial phase-space distribution. The space-charge 
radial electric field of an axisymmetric beam can be found 
from Gauss' law by counting the numbers of particles in 
cells of a finite radial grid, which extends up to four times the 
beam matched radius. The initial phase-space state is populated 
randomly but in accordance with (\ref{fn}) for a chosen $n$.
The matched distributions remain stable except for a minor dilution 
related to numerical errors. However, even the matched KV beam is 
unstable for $\eta \le 0.4$, in agreement with existing theory 
\cite{RLG70} and earlier simulations \cite{HLSH}. 

The beam breathing oscillations are excited by loading a mismatched 
initial distribution $r_i = \mu \tilde{r}_i$, 
$r'_i = \tilde{r}'_i/\mu $, where $\tilde{r}_i,\tilde{r}'_i$ 
correspond to the matched one, and the mismatch parameter $\mu \le 1$. 
A typical range of the simulation parameters: time step 
$\Delta t = T/100$, where $T$ is the period of breathing oscillations,
total number of particles $N_{par}=16K\mbox{ to }4096K$, where $K=1024$, 
and radial mesh size $\Delta r = a/128\mbox{ to }a/16$.
The code performs simulations of about 100 breathing 
oscillations per CPU hour for $N_{par}=256K$ on Sun UltraSparc 1/170. 

The beam behavior is studied as a function of tune depression $\eta$ 
and mismatch $\mu$. Due to a discrete filling of a mismatched 
beam distribution in simulations and, for $n \ge 1$, due to non-linear
space-charge forces, higher modes are excited in addition to the 
breathing mode. Some of them can be unstable depending on values of 
$\eta$ and $\mu$. A detailed numerical study of stability and halo 
formation for the KV beam and its comparison with the theory 
predictions \cite{RLG70,RLG94} have already been reported 
in \cite{GCY,GCKY}. 
Here we compare results for different transverse distributions. 
In Figs.~1-2 the maximal radius of the whole ensemble of particles 
is plotted versus the number of breathing oscillations for the KV 
and WB beams, for the particular case of $\eta=0.7$ and
$\mu=0.8$ ($N_{par}=256K$, $\Delta r=a/64$). 

\begin{figure}[htb]
\centerline{\epsfig{figure=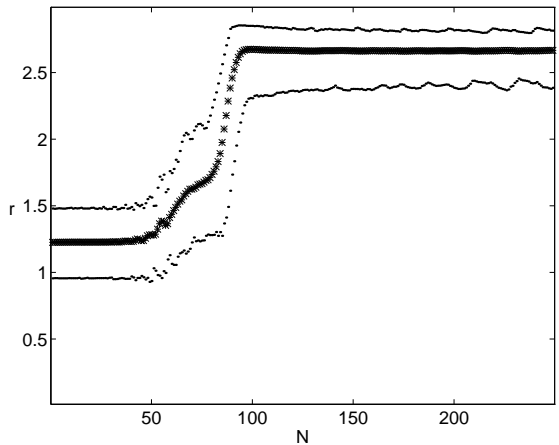,width=7.5cm}}
\caption{KV beam radius versus the number of breathing periods 
for $\eta=0.7$ and $\mu=0.8$. Stars are for period averages, 
dots show minimum and maximum during a period.}
\end{figure}

\begin{figure}[htb]
\centerline{\epsfig{figure=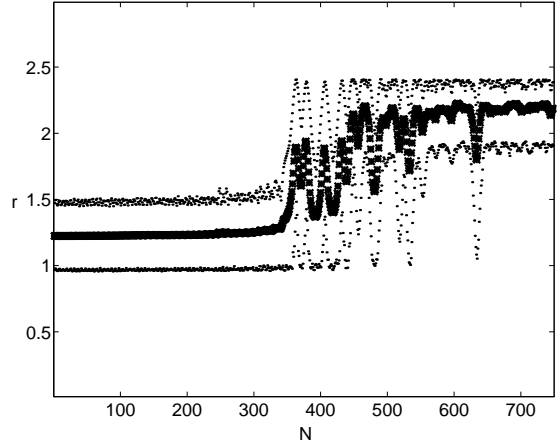,width=7.5cm}}
\caption{The same as Fig.~1, but for the WB beam.}
\end{figure}

Comparison of Figs.~1 and 2 shows that for these  parameters the 
WB beam remains stable much longer than the KV one, but eventually
it also blows up and some particles form a halo far from the beam 
core. Results for the $n=2$ distribution are similar to those for WB.
Some results depend on simulation parameters; e.g., it takes a
smaller number of the breathing periods for a beam to blow up
if $N_{par}$ is smaller (i.e., higher noise). However, the 
maximum radius, as well as the fraction of particles outside the
core, are practically independent of $N_{par}$. The number of 
particles which go into the halo and produce jumps of $r_{max}$ seen
in Figs.~1-2, might be rather small. We define the halo intensity 
$h$ as the number of particles outside the boundary $r_{b}=1.75 a$ 
divided by $N_{par}$.
Such a definition is arbitrary, but convenient to compare beam halos 
over a wide range of tune depressions. 
While the beam behavior in Figs.~1 and 2 seems qualitatively similar, 
the halos for these two cases are very different: 
$h \simeq 3.5\cdot10^{-3}$ for KV, and about 100 times less for the
WB, with only a few particles in the halo (less than 10 of 256K).
That is the reason for oscillations of $r_{max}$ in Fig.~2: these 
few halo particles can initially all come back to the core 
simultaneously.

\begin{figure}[htb]
\centerline{\epsfig{figure=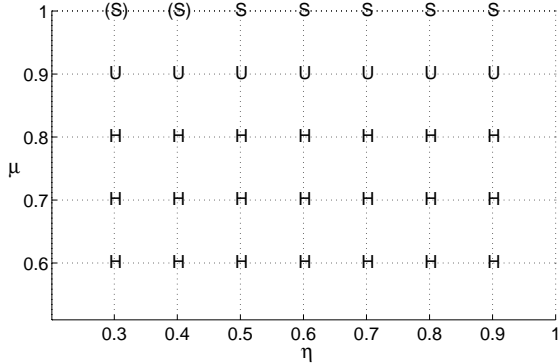,width=7.5cm}}
\caption{Beam behavior (qualitatively) versus tune depression $\eta$
and mismatch $\mu$.}
\end{figure}

A qualitative picture of the beam behavior for various values of the 
tune depression and mismatch is shown in Fig.~3, and is practically 
the same for all distributions studied.
'H' corresponds to beam instability with halo formation,
usually with a noticeable emittance growth, 'U' means that the beam 
is unstable but a halo is not observed in our simulations, 
and 'S' indicates beam stability. 
The most surprising feature of the diagram is the lack of any 
significant dependence on $\eta$ for mismatched beams; on the 
contrary, the qualitative changes depend primarily on $\mu$. 
When $\mu$ changes from 0.6 to 0.8, the ratio 
$\tilde{\mathcal{E}}_{fin} /\tilde{\mathcal{E}}_{ini}$
decreases from 1.7--2 to 1.03--1.07 
for the KV beam, and from 1.4--1.5 to 1.00--1.01 for the WB and $n=2$.
The number of breathing periods after which the beam radius starts 
to grow noticeably and the halo forms, has some dependence on $\eta$; 
it is smaller for small $\eta$. 

We performed a systematic study of the KV, WB, and $n=2$ distributions 
for tunes $\eta$ from 0.1 to 0.9 and mismatches $\mu$ from 0.6 to 1.0.
Figure~4 shows the ratios of the halo radius to that of the matched
beam for the KV and WB beams with three different mismatches,
$\mu=0.6$, 0.7, and 0.8. Results for $n=2$ beam are not shown; 
they are slightly lower than those for the WB beam. The KV halo has 
a larger radius, especially with small space charge 
(large $\eta$), but for space-charge dominated beams, at very small 
$\eta$, the ratios converge for all distributions. The analytical
model for the KV halo formation \cite{RLG94} predicts finite values 
of $r_{max}/a$ between 2 and 2.5 depending on $\eta$ and $\mu$. 
One can see from simulations that it works well also for WB and 
$n=2$ beams. 

\begin{figure}[htb]
\centerline{\epsfig{figure=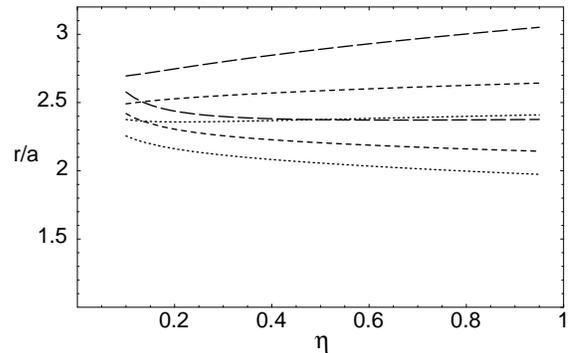,width=7.5cm}}
\caption{Ratio of halo radius to that of the matched beam for KV 
(top curves) and WB (bottom curves) beams versus tune depression 
$\eta$ for different mismatches: $\mu=0.6$ long-dashed, $\mu=0.7$ 
short-dashed, $\mu=0.8$ dotted.}
\end{figure}

Simulation results for halo intensity $h$ are shown in Fig.~5 for KV
and WB distributions. Again, results for $n=2$ are just slightly lower
than for the WB beam, and not shown. The intensity depends essentially
on the mismatch, and decreases quickly as the mismatch decreases.
The WB halo is about 2--3 times less intense than the KV halo 
for small space charge and large mismatch (0.6 and 0.7) but, for 
space-charge dominated beams, the intensities are about the same.
For $\mu=0.8$, however, the WB halo is at least an order of 
magnitude less intense than the KV one;  it is not even
included in Fig.~5. An apparent decrease of $h$ as $\eta$ 
decreases is due to the definition used: the halo boundary radius
$r_b=1.75a$ increases as $1/\sqrt{\eta}$. If a fixed boundary is 
used instead, the same for all tunes, the halo intensity would be 
larger for larger space charge.

\begin{figure}[htb]
\centerline{\epsfig{figure=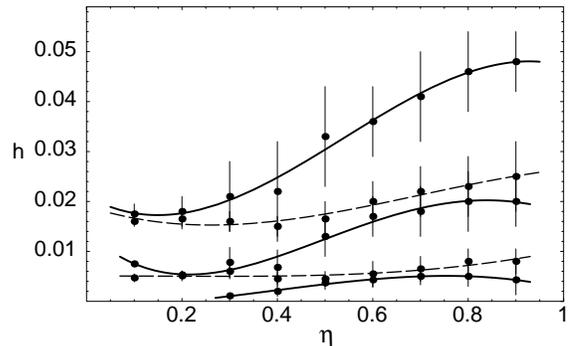,width=7.5cm}}
\caption{Halo intensity for KV (solid) and WB beams (dashed)
vs tune depression $\eta$ for mismatches $\mu=0.6$ 
(top pair), $\mu=0.7$ (middle pair), $\mu=0.8$ (bottom, KV only).}
\end{figure}

One more interesting feature is how fast the halo develops. For
the KV beam, the process is usually rather fast, and the halo 
saturates after a few hundred breathing periods. For the WB and 
$n=2$ distributions, it continues to grow rather slowly, and 
asymptotic values are usually reached after a few thousand 
breathing oscillations; it takes especially long for 
$\eta \le 0.3$. Data plotted in Fig.~5
correspond to the asymptotic values, after N=5000 breathing 
periods for WB and after N=600 for KV (except $\eta \le 0.2$,
where KV results are also for N=5000). These 5000 breathing
oscillations correspond to 5--10~km of the length for a typical
machine, much longer than existing proton linacs.  

\section{Conclusions}

Our simulations show the qualitative similarity of the beam behavior
for all transverse distributions studied. The KV beam can be 
considered as an extreme case compared to the WB and $n=2$ 
distributions which are closer to real beams. The halo intensity
is a few times higher and saturates faster for the KV 
distribution than for the other two.

An interesting new observation is that for axisymmetric beams 
under consideration the beam stability and halo formation depend 
primarily on the mismatch, not on the tune shift. The halo was clearly 
observed only for large mismatches, at least 20\%, and its radius 
is in agreement with the analytical model \cite{RLG94} for halo 
formation.


\begin{thebibliography}{9}

\bibitem{RLG94}
 R.L.~Gluckstern, Phys.\ Rev.\ Letters {\bf 73}, 1247 (1994).
\bibitem{KV} 
I.M.~Kapchinsky and V.V.~Vladimirsky, in {\it Pro\-ceed.\ Int.\ 
Conf.\ on High Energy Accelerators} (CERN, Geneva, 1959), p.~274.
\bibitem{Okam} 
H.~Okamoto and M.~Ikegami, Phys.\ Rev.\ E {\bf 55}, 4694 (1997).
\bibitem{MR} 
M.~Reiser, {\it Theory and Design of Charged Particle Beams}
 (Wiley, New York, 1993).
\bibitem{RLG70} 
R.L.~Gluckstern, 
in {\it Proceed.\ of the Linac Conference} (Fermilab, 1970), p.~811.
\bibitem{HLSH} 
I. Hofmann, L.J. Laslett, L. Smith and I. Haber, Particle
Accelerators {\bf 13}, 145 (1983).
\bibitem{GCY} 
R.L.~Gluckstern, W.-H.~Cheng and H.~Ye, 
Phys.\ Rev.\ Letters {\bf 75}, 2835 (1995).
\bibitem{GCKY} 
R.L.~Gluckstern, W.-H.~Cheng, S.S.~Kurennoy and H.~Ye, 
Phys.\ Rev.\ E {\bf 54}, 6788 (1996).
\end{thebibliography}
\end{document}